\documentclass[runningheads,a4paper]{llncs}

\usepackage{hyperref}
\usepackage{tikz}
\usetikzlibrary{decorations.markings}
\usetikzlibrary{arrows}
\usepackage{amssymb}
\usepackage{amsfonts,amsmath,latexsym,stmaryrd}
\usepackage{cite}
\usepackage{theorem}
\usepackage{wrapfig}
\usepackage{alltt}
\usepackage{fancyvrb}
\usepackage{relsize}
\usepackage{multirow} 
\usepackage{hhline} 
\usepackage{listings}
\theorembodyfont{\rmfamily}

\begin{document}

\title{Proof-Pattern Recognition and Lemma Discovery in ACL2\thanks{The work was supported
    by EPSRC grants EP/J014222/1$^{1}$, EP/H024204/1$^{3}$ and a VINNMER Marie Curie Fellowship$^2$.}}

\author{J\'onathan Heras\inst{1} \and Ekaterina Komendantskaya\inst{1} \and Moa Johansson\inst{2} \and Ewen Maclean\inst{3}}
\authorrunning{J. Heras,  E. Komendantskaya, M. Johansson and E. Maclean}

\institute{School of Computing, University of Dundee, UK 
\and Dept. of Computer Science and Engineering, Chalmers University, Sweden
\and School of Informatics, University of Edinburgh, UK
\email{\{jonathanheras,katya\}@computing.dundee.ac.uk, moa.johansson@chalmers.se, E.Maclean@ed.ac.uk}}

\maketitle

\begin{abstract}
We present a novel technique for combining statistical machine learning for proof-pattern recognition with symbolic methods for lemma discovery. The resulting tool, \textbf{ACL2(ml)}, gathers proof statistics and uses statistical pattern-recognition to pre-processes data from libraries, and then suggests auxiliary lemmas in new proofs by analogy with already seen examples. This paper presents the implementation of ACL2(ml) alongside theoretical descriptions of the proof-pattern recognition and lemma discovery methods involved in it.   



\end{abstract}

\section{Introduction}\label{sec:intro}


Over the last few decades, theorem proving has seen major developments.
\emph{Automated (first-order) theorem provers (ATPs)} (e.g. E, Vampire, SPASS) and SAT/SMT solvers
(e.g. CVC3,  Yices, Z3) are becoming increasingly fast and efficient. \emph{Interactive (higher-order) theorem provers (ITPs)} (e.g. Coq,
Isabelle/HOL, Agda, Mizar) have been enriched with dependent types, 
(co)inductive types, type classes and provide rich programming environments.

The main conceptual difference between  ATPs and ITPs lies in the styles of proof development. 
For ATPs, the proof process is primarily an automatically performed \emph{proof search} in a \emph{first-order} language. 
In ITPs, the proof is guided by \emph{the user} who specifies which tactics to apply.
ITPs often work with  \emph{higher-order} logic and type theory. 

Communities working on development, implementation and applications of ATPs and ITPs have
accumulated big corpora of electronic proof libraries. However, the size of the libraries, as well as 
their technical and notational sophistication often stand in the way of efficient knowledge re-use.
Very often, it is easier to start a new library from scratch rather than search the existing proof libraries
for potentially common heuristics and techniques.

Pattern-recognition \cite{Bishop} is an area of machine-learning that develops statistical methods for discovering patterns in data.
In the statistical sense, a pattern is a correlation of several numeric features, by which the data is represented. 
 For instance, the correlation between the theorem statement and the auxiliary lemmas required in its proof can be data-mined to improve automation of premise selection
  in ATPs~\cite{DFGS99,DenzingerS00,KuhlweinLTUH12,TsivtsivadzeUGH11,UrbanSPV08}.
The history of successful and unsuccessful proof attempts and proof steps can also be used to inform interactive proof development in
  ITPs~\cite{Duncan02,KHG12}.

Statistical machine-learning methods 
are well-suited for fast processing of big proof libraries, adapt well to proofs of varied sizes and complexities (first- or higher-order);
and are generally tolerant to noise. However, 
they have very weak capacities for conceptualisation.
For instance, ML4PG~\cite{KHG12} only displays the families of related proofs to the Coq user, and can tell the correlation of which features formed the pattern, but it neither explains why this happens nor formulates any conceptual proof hints. 

While statistical methods focus on extracting information from existing large theory libraries, many symbolic methods are instead concerned with automating the discovery of lemmas in new theories~\cite{HR,mathsaid,JDB11,montano2012,hipspec,HLW12},
while relying on existing proof strategies, e.g. proof-planning and rippling \cite{BB05}. 
These systems are naturally more deterministic and  algorithmic than statistical AI.  

IsaCoSy~\cite{JDB11} and IsaScheme~\cite{montano2012} are two term synthesis systems built on top of Isabelle.
They generate candidate conjectures which are filtered through a counter-example checker. Surviving conjectures are passed 
to an automated prover and those proven are added to the theory. The systems differ in their heuristics for term generation: IsaCoSy only generates irreducible terms, while IsaScheme uses \emph{schemes}
\cite{Buchberger04} to specify the shapes of candidate theorems. A similar system is QuickSpec and its extension HipSpec \cite{quickspec,hipspec}, which generates equational theorems about Haskell programs, using congruence closure for conjecture generation. 
MATHsAiD~\cite{mathsaid} generates theories by forward reasoning from a given set of
axioms, applying various `interestingness' heuristics. 

Symbolic methods have limits: they can be slow on large inputs due to the increase in the search space, rely on having access to good counter-example finders for filtering of candidate conjectures and require existing proof strategies for proving those remaining. 

\begin{figure}[t]
\centering 
\footnotesize{
\begin{tikzpicture}

\draw[fill=black] (-1,2) -- (0,2) -- (0,2.2) -- (-1,2.1) --cycle;
\draw[fill=black] (.5,2) -- (1.5,2) -- (1.2,2.2) -- (1.2,3) -- (.7,3) -- (.7,3.5) -- (.6,3.5) -- (.6,2.3) -- (.7,2.3) -- (.7,2.9) -- (1.1,2.9) -- (1.1,2.2) -- cycle;

\draw[-latex,shorten <=2pt,shorten >=2pt,dashed] (1.2,2.5) -- (6,2.5); 
\draw (-1,3) node{{\scriptsize User}};

\draw[latex-,shorten <=2pt,shorten >=2pt,dashed] (0,2) -- (0,0); 
\draw[-latex,shorten <=2pt,shorten >=2pt,dashed] (8,3) -- (8,.8); 
\draw (8,1.6) node[anchor=north,fill=white]{\emph{\scriptsize{feature extraction}}};
\draw (0,1.5) node[anchor=north,fill=white]{\emph{\scriptsize{new lemmas}}};
\draw[fill=white] (6,2) rectangle (10,3); 
\draw  (8,2.5) node{{\scriptsize Interface of ACL2(ml)}};

\draw[fill=gray,draw=gray] (-1.95,.75) rectangle (2.05,-.25);  
\draw[fill=white] (-2,.8) rectangle (2,-.2); 
\draw  (0,0.5) node{{\scriptsize Symbolic Lemma Discovery}}; 
\draw (0,0.1) node{{\scriptsize (Lemma Analogy)}};

\draw[fill=gray,draw=gray] (5.9,.75) rectangle (10.2,-.25);  
\draw[fill=white] (5.85,.8) rectangle (10.15,-.2); 
\draw  (8,0.5) node{{\scriptsize Statistical Machine-Learning}}; 
\draw (8,0.1) node{{\scriptsize (Clustering)}};

\draw[latex-,shorten <=2pt,shorten >=2pt,dashed] (1.5,2) -- (5.9,.4);
\draw (4,1.47) node[anchor=north,fill=white]{\emph{\scriptsize{similar lemmas}}};
\draw[latex-,shorten <=2pt,shorten >=2pt,dashed] (2,0.25) -- (5.9,0.25); 
\draw (4,0.47) node[anchor=north,fill=white]{\emph{\scriptsize{similar lemmas}}};
 

%
%

\end{tikzpicture}}
 \caption{{\scriptsize \emph{Architecture of proof-pattern recognition methods in ACL2(ml)}. The Emacs interface for ACL2 extracts 
 important features from ACL2 theorems, connects to machine-learning software, clusters the theorems there, and sends the result
 (families of similar theorems) to the screen. In addition, for an unproved theorem $T$, it sends the cluster $C$ of $T$ 
 to the Lemma Analogy tool, which in its turn generates an auxiliary lemma by analogy with auxiliary lemmas of the theorems in $C$.  }}\label{fig:interaction}
\end{figure}
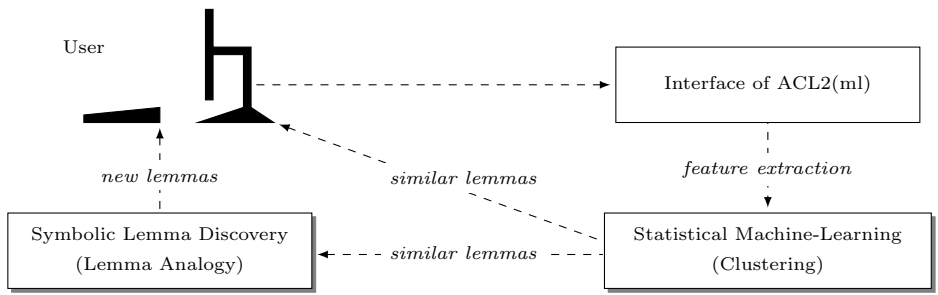

In this paper, we show that it is possible to combine statistical and symbolic methods to get the best of both worlds: statistical pattern-recognition methods are well-suited for finding families of similar proofs; symbolic tools can use 
this data for more efficient lemma discovery. Feeding outputs of
one algorithm to the other leads to a very natural proof-pattern recognition system, see Figure~\ref{fig:interaction}.
To realise this general thesis about the synthesis of the two styles of proof-pattern recognition, we have made the following methodological choices:

\textbf{1)} We chose the ACL2 prover for our experiments: it is based on first-order logic and has features of both ITPs and ATPs. ACL2 will try to prove given conjectures automatically, but strongly relies on the user having supplied the auxiliary lemmas required for rewriting. Thus, the user must often intervene and advance the proof by adding new lemmas, much like in ITPs.
Section \ref{sec:ACL2} gives a brief introduction to ACL2.


\textbf{2)} On the machine-learning side, we chose to adapt the ML4PG\cite{KHG12} design of \emph{interactive} proof-pattern recognition; cf. Figure~\ref{fig:interaction}. We extended this work with a new feature extraction algorithm for ACL2(ml), based on term trees and recurrent clustering. 
Much like in ML4PG, we discover families of related lemmas, rather than 
premise hierarchies like in \cite{KuhlweinLTUH12,TsivtsivadzeUGH11,UrbanSPV08}.
Unlike  \cite{KuhlweinLTUH12,TsivtsivadzeUGH11,UrbanSPV08}, we use clustering instead of supervised learning; and do not use sparse methods. Section~\ref{sec:acl2ml} explains our new method in detail.

\textbf{3)} On the symbolic side, we develop a novel lemma generation approach
  which uses the statistical suggestions to reduce the
  search space. 
	Speculating the correct lemma in a proof can often be the
  \emph{eureka} moment which allows a proof to succeed. It is these
  eureka lemmas which we would like to reproduce automatically. 
	We propose an analogy driven approach
  where the term structure of an example \emph{source lemma} is analysed
  to produce an analogous \emph{target lemma} for a given target
  problem. In the context of ITP, the user is then presented with an auxiliary lemma suggestion which is customised to the particular problem on
  which he is working. Section~\ref{sec:lemma-analogy} is devoted to this subject. 


We implement the above choices in the ACL2 extension ACL2(ml), available at~\cite{acl2ml}.
Finally, we conclude with evaluation of this method in Section~\ref{sec:concl}.
The results we obtain with ACL2 have a wider significance, as  methodology we develop here would apply to a wide range of first-order theorem provers.




 


\section{Background}\label{sec:ACL2}

 ACL2~\cite{kaufmann00acl2,acl2-web} (standing for \emph{A Computational Logic for an Applicative Common Lisp}) 
 is a programming language, a logic, and
 a theorem prover supporting reasoning in the logic. The ACL2 programming
 language is an extension of an applicative subset of Common Lisp. The ACL2
 logic is an untyped first-order logic with equality, used for specifying properties
 and reasoning about the functions defined in the programming language. All 
 the variables in the formulas allowed by the ACL2 system are implicitly 
 universally quantified. The syntax of its
 terms and formulas is that of Common Lisp.


 \begin{example}\label{ex:example0}
 Given the recursive and tail-recursive functions to compute factorial, $2^n$ and Fibonacci, 
 the ACL2 user can specify the equivalence between the functions as shown in Figure~\ref{fig:rec-tail} with theorems 
 \verb"fact-fact-tail", \verb"power-power-tail" and \verb"fib-fib-tail". Note the similarity between these lemmas.
 
 \end{example}

 \begin{figure}[t]
  
\begin{Verbatim}[frame=single,fontsize=\scriptsize,commandchars=\\\{\}]
;; \textit{Factorial}
1 (defun fact (n) (if (zp n) 1 (* n (fact (- n 1)))))

2 (defun helper-fact (n a) (if (zp n) a (helper-fact (- n 1) (* a n))))

3 (defun fact-tail (n) (helper-fact n 1))

4 (defthm fact-fact-tail (implies (natp n) (equal (fact-tail n) (fact n))))

;; \textit{2^n}
1 (defun power (n) (if (zp n) 1 (* 2 (power (- n 1)))))

2 (defun helper-power (n a) (if (zp n) a (helper-power (- n 1) (+ a a))))

3 (defun power-tail (n) (helper-power n 1))

4 (defthm power-power-tail (implies (natp n) (equal (power-tail n) (power n))))

;; \textit{Fibonacci}
1 (defun fib (n) (if (zp n) 0 (if (equal n 1) 1 (+ (fib (- n 1)) (fib (- n 2))))))

2 (defun helper-fib (n j k) (if (zp n) j (if (equal n 1) k (helper-fib (- n 1) k (+ j k)))))

3 (defun fib-tail (n) (helper-fib n 0 1))

4 (defthm fib-fib-tail (implies (natp n) (equal (fib-tail n) (fib n)))) 
\end{Verbatim}
\vspace{-.5cm}
 \caption{{\scriptsize \emph{ACL2 definitions and theorems.} 1: recursive arithmetic functions.  2: helpers of tail-recursive arithmetic functions.
 3: tail-recursive arithmetic functions. 4: Equivalence theorems of recursive and tail-recursive functions.}}\label{fig:rec-tail}
\end{figure}


 
 ACL2 has both automatic and interactive features. 
 It is automatic in the sense that once a \texttt{defthm} command is submitted,
 the user can no longer interact with the system (the ACL2 proof engine applies a collection of automatic tactics until either the conjecture is proven or none of the tactics is applicable).
 Often, non-trivial results cannot be proven on the first attempt. The user
then has to interact with the prover by supplying a suitable
 collection of definitions and auxiliary lemmas, used in subsequent proofs as rewriting rules.
 These lemmas are suggested by a preconceived hand proof or by inspection of
 failed proofs; this kind of interaction is referred to as \emph{The Method}~\cite{kaufmann00acl2}.

  \begin{example}\label{example_sl}
 ACL2's first attempt at proving the conjecture \verb"fact-fact-tail" of Example~\ref{ex:example0} fails, and the user needs to introduce a lemma (inspired by the failed proof of \verb"fact-fact-tail"), which is automatically proven by ACL2:  
   \begin{Verbatim}[fontsize=\footnotesize]
(defthm fact-fact-tail-helper (implies (and (natp n) (natp a)) 
       (equal (helper-fact n a) (* a (fact n)))))
 \end{Verbatim}
 ACL2 can now automatically prove the conjecture \verb"fact-fact-tail", using this auxiliary result. In later sections, we will show how ACL2(ml) automatically detects  similar theorems 
 and generates auxiliary lemmas for them.
\end{example}

The proofs of equivalence between the recursive and tail-recursive functions in Figure~\ref{fig:rec-tail} follow a common pattern: the equivalence theorems are not proven by ACL2 in the first attempt, and the user must introduce auxiliary lemmas about the \verb"helper" functions.
In general, the detection of common patterns and the generation of new lemmas from those patterns is based on user's experience and 
can be challenging.
The ACL2 distribution consists of several libraries containing hundreds of theorems, developed by several users with their own notations.  It can be a challenge to detect patterns across different users, notations and libraries.
Moreover, given two similar theorems $T_1$ and $T_2$, the lemmas used to prove $T_1$ can be substantially different to the ones needed for $T_2$ (e.g. different function symbols, different lemma structures, additional conditions, new concepts). 
 

In the next sections, we describe how these challenges can be addressed automatically. 
We consider two running examples. The first one is the JVM library developed in~\cite{Moore03provingtheorems}.  
This library contains the correctness proofs of the Java bytecode associated with several arithmetic programs such as multiplication, 
factorial and Fibonacci, including the theorems of Figure~\ref{fig:rec-tail}. 
As a second running example, we consider the \emph{Lists} library presented in~\cite{KM-prove}. From this library, we consider three functions that will appear later in the paper: \verb"sort" 
(that sorts a list of natural numbers), \verb"rev" (which reverses a list), and \verb"int" (which takes two lists as arguments and returns the list of elements
that appear in both); and three theorems about these functions.
 
 \begin{Verbatim}[fontsize=\footnotesize]
(defthm sortsort (implies (nat-listp x) (equal (sort (sort x)) (sort x))))
(defthm revrev (implies (true-listp x) (equal (rev (rev x)) x)))
(defthm int-x-x (implies (true-listp x) (equal (int x x) x)))
 \end{Verbatim}

\section{Statistical Proof-Pattern Recognition with ACL2(ml)}\label{sec:acl2ml}


We present a statistical proof-pattern recognition extension to ACL2, called \textbf{ACL2(ml)}. 
Its implementation design follows the ML4PG tool for Coq \cite{KHG12}: this Emacs-based machine-learning extension works on the background
of the theorem prover, gathers statistics of proof features, and then, on user's request, connects to a machine-learning toolbox (MATLAB or Weka); groups the proofs 
using \emph{clustering} algorithms, 
and displays families of related proofs to the user.
This approach allows real-time interaction between the user, the prover, and the machine-learning systems; as \cite{KHG12} explains in detail.

Here, we concentrate only on original components of ACL2(ml) as compared to ML4PG.  
There are two such features: a novel algorithm for term tree feature extraction and the method of \emph{recurrent clustering}.

\textbf{\emph{Term tree feature} extraction in ACL2(ml).} 
The discovery of statistically significant features in data is a research area of its own in machine
learning, known as \emph{feature extraction}~\cite{Bishop}. Irrespective of the particular algorithm
used, most pattern-recognition tools will require that the features have numeric values; and the number of selected features is limited and fixed 
(sparse methods, like the ones applied in e.g.~\cite{KuhlweinLTUH12,TsivtsivadzeUGH11,UrbanSPV08}, are the exception to the latter rule).
The feature extraction algorithm implemented in ML4PG was based on correlation between the goal shapes, the tactics and tactic parameters within a few steps of the interactive proof.
As ACL2 is not a tactic-driven language, we need to extract features from the ACL2 terms directly. Feature extraction from terms or \emph{term trees} is common to most feature-extraction algorithms implemented in automated theorem provers: see e.g.~\cite{KuhlweinLTUH12,TsivtsivadzeUGH11,UrbanSPV08}.

\begin{definition}[Term tree]
A variable or a constant is represented by a tree consisting of one single node, labelled by the variable or the constant itself.
A function application $f(t_1, \ldots,  t_n)$ is represented by the tree with the root node labelled by $f$, and its immediate subtrees given by trees representing $t_1 , \ldots, t_n$.
\label{def:termtree}
\end{definition}

\begin{figure}[t]
 \centering
 
 \begin{tikzpicture}
  \draw (0,-0.25) -- (-1.5,-1) -- (-1.5,-1.75); 
  \draw (0,-0.25) -- (1.5,-1) -- (0.5,-1.75) -- (0.5,-2.5); 
  \draw (0,-0.25) -- (1.5,-1) -- (2.5,-1.75) -- (2.5,-2.5); 
  \draw (0,-0.25) node[fill=white]{{\scriptsize \verb"implies"}};
  \draw (-1.5,-1) node[fill=white]{{\scriptsize \verb"natp"}};
  \draw (-1.5,-1.75) node[fill=white]{{\scriptsize \verb"n"}};
  \draw (1.5,-1) node[fill=white]{{\scriptsize \verb"equal"}};
  \draw (0.5,-1.75) node[fill=white]{{\scriptsize \verb"fact-tail"}};
  \draw (2.5,-1.75) node[fill=white]{{\scriptsize \verb"fact"}};
  \draw (0.5,-2.5) node[fill=white]{{\scriptsize \verb"n"}};
  \draw (2.5,-2.5) node[fill=white]{{\scriptsize \verb"n"}};

	\draw (6.5,-1.5) node{{\scriptsize 
 \begin{tabular}{|c||c|c|c|c|c|c|c|c|}
     \hline
    & impl & natp & equal & fact-t. & fact & n & n' & n''  \\
    \hline
    \hline
    impl & 0 & 1 & 1  & 0  & 0 & 0 & 0 & 0 \\
    \hline
    natp & 0 & 0 & 0 & 0 & 0 & 1 & 0 & 0  \\
    \hline
    equal & 0 & 0 & 0 & 1 & 1  & 0 & 0 & 0 \\
    \hline
    fact-t. & 0 & 0 & 0 & 0 & 0  & 0 & 1 & 0 \\
		 \hline
    fact & 0 & 0 & 0 & 0 & 0  & 0 & 0 & 1 \\
    \hline
    n &  0 & 0 & 0 & 0 & 0  & 0 & 0 & 0 \\
    \hline
    n' & 0 & 0 & 0 & 0 & 0  & 0 & 0 & 0 \\
      \hline
			 n'' & 0 & 0 & 0 & 0 & 0  & 0 & 0 & 0 \\
      \hline
 \end{tabular}}
	
};
 \end{tikzpicture}
 \caption{{\scriptsize \emph{Term tree and adjacency matrix for Theorem \texttt{fact-fact-tail}; assuming the root-to-leaf direction of edges.}}}\label{fig:tree}
\end{figure}
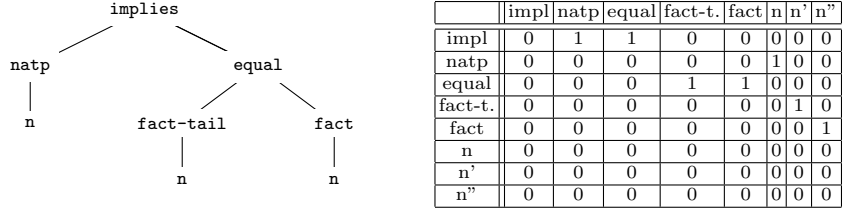

\begin{example}
 Theorem \verb"fact-fact-tail" from Example~\ref{ex:example0} can be represented by the term tree of Figure~\ref{fig:tree}. 
\end{example}

A variety of methods exists to represent trees as matrices, for instance using \emph{adjacency matrices} and \emph{incidence matrices}, with the former shown in Figure \ref{fig:tree}.
The adjacency matrix and the various previous methods of term tree feature extraction (e.g. \cite{KuhlweinLTUH12,TsivtsivadzeUGH11,UrbanSPV08}) share the following common 
properties: different tree nodes are represented by distinct matrix entries (\emph{features}); 
the matrix entries (\emph{feature values}) are binary; and the size of the matrix depend on the tree size.
For big libraries, such matrices can grow very large (up to $10^6$ in some experiments) 
and at the same time very sparse, which implies the use of \emph{sparse} machine-learning in  \cite{KuhlweinLTUH12,TsivtsivadzeUGH11,UrbanSPV08}.

In ACL2(ml), we implement a more compact feature extraction, where a total number of features is fixed for all libraries at $49$, and the average density of matrices is
$20\%$. 
The smaller and denser 
feature vectors can be used to data-mine  both big and small data sets. 
Large and sparse feature vectors would require to use big data sets of proofs, where the size of the example data set is comparable to the number of features. 
In our interactive setting, it is crucial that the ACL2(ml) tool works equally well with both big and small proof libraries; and that the user can interact with it at any stage of the proof development.

We develop a method to overcome the problem of having to track a large (potentially unlimited)
number of ACL2 symbols by a finite number of features, as follows.
The ACL2 symbols are represented by distinct feature values given by rational numbers. 
The feature values are computed dynamically by \emph{recurrent clustering} algorithm, thus reflecting the recursive nature of the ACL2 functions and proofs. 
The \emph{features} are given by the finite number of properties common to all possible term trees: the term arity and the term tree depth; 
see Table \ref{tab:tt}. This is formalised in the following definitions.

 
\begin{definition}[Term tree depth level]
Given  a term tree $T$, the \emph{depth} of the node $t$ in $T$, denoted by \emph{depth(t)}, is defined as follows:

$-$ $depth(t) = 0$, if $t$ is a root node;

$-$ $depth(t) = n+1$, where $n$ is the depth of the parent node of $t$.
\label{def:termtreelevel}
\end{definition}
	
\begin{definition}[ACL2(ml) term tree matrices]\label{df:matrix}
Given a term tree $T$ for a term with signature $\Sigma$, and a function $[.]: \Sigma \rightarrow \mathbb{Q}$, the ACL2(ml) term tree matrix $M_T$ is a $7 \times 7$ matrix that satisfies the following conditions:

$-$ the $(0,j)$-th entry of $M_T$ is a number $[t]$, such that $t$ is a node in $T$, $t$ is a variable and $depth(t) = j$.

$-$ the $(i,j)$-th entry of $M_T$ ($i \neq 0$) is a number $[t]$, such that $t$ is a node in $T$, $t$ has arity $i+1$ and $depth(t) = j$. 
\end{definition}

We deliberately specify $[.]$ only by its type in Definition \ref{df:matrix}. In ACL2(ml), this function 
is dynamically re-defined for every library and every given proof stage, as we are going to describe shortly in Definition~\ref{def:recurrent}. 
In practice, there will be a set of such functions computed in every session of ACL2(ml). 


\begin{table}[t]

\begin{tikzpicture}

\draw (0,0) node{
\scriptsize{
 \begin{tabular}{|c||c|c|c|c|c|}
     \hline
    & variables & arity 0 & arity 1 & arity 2  \\
    \hline
    \hline
    $td0$ & 0 & 0 & 0  & [implies]  \\
    \hline
    $td1$ & 0 & 0 & [natp] & [equal]  \\
    \hline
    $td2$ & [n] & 0  & [fact-tail]::[fact] & 0  \\
    \hline
    $td3$ & [n]::[n]& 0 & 0 & 0 \\
    \hline
 \end{tabular}}};

\draw (6,0) node{

\scriptsize{
 \begin{tabular}{|c||c|c|c|c|c|c|}
     \hline
    & variables & arity 0 & arity 1 & arity 2 & arity 3 \\
    \hline
    \hline
    $td0$ & 0 & 0 & 0  & 0  & [if] \\
    \hline
    $td1$ & 0 & [1] & [zp] & [*] & 0 \\
    \hline
    $td2$ & [n]::[n] & 0  & [fact] & 0 & 0  \\
    \hline
    $td3$ & 0& 0 & 0 & [-] & 0\\
    \hline
    $td4$ & [n]& [1] & 0 & 0 & 0\\
    \hline
 \end{tabular}}};
 \end{tikzpicture}

  \caption{\scriptsize{\emph{\textbf{Left.} Dense term tree feature extraction matrix for theorem \texttt{fact-fact-tail}.
  \textbf{Right.} Dense term tree feature extraction matrix for definition \texttt{fact}.} The operator ``\texttt{::}'' indicates the concatenation of values (e.g.
  4::5=45).}}\label{tab:tt}
	\vspace{.5cm}
 \end{table}

%

	
	To make the feature extraction uniform across all ACL2 terms appearing in the library, the matrices are extended to cover
	terms up to arity $n$ and tree-depth $m$.
	The parameters $n$ and $m$ can vary slightly; for all libraries considered in the paper $n=5$ and $m=7$ were sufficient, giving a feature vector size of $49$ 
	-- a small size compared to sizes up to $10^6$ in sparse approaches \cite{KuhlweinLTUH12,TsivtsivadzeUGH11,UrbanSPV08}.
	Having a clustering algorithm working for small sets of examples is crucial for the technique of recurrent clustering we implement for ACL2(ml). 

\textbf{Recurrent Clustering.}
As discussed above, the function $[.]$ 
influences results of ACL2(ml) proof-pattern recognition; 
and the computation of $[.]$ needs to be sensitive to the similarities that exist between the symbols appearing in the proofs. 
As ACL2 is a functional language, every entry in a term
tree is necessarily itself defined in ACL2. This symbol definition can itself be clustered against other definitions used by the
library; and 
the process can be repeated recursively to include all the necessary 
standard library definitions. This is how the feature extraction becomes a part of \textbf{recurrent clustering} in ACL2(ml).

We first define some essential clustering parameters. ACL2(ml) connects automatically to clustering~\cite{Bishop} 
algorithms available in Weka (K-means, FarthestFirst and E.M.). 
Clustering techniques divide data into $n$ groups of similar objects (\emph{clusters}), where the value
of $n$ is a learning parameter provided by the user. Increasing the value of $n$ makes  the algorithm separate objects into more classes, and, as a consequence, each cluster will contain fewer examples. 
There is a number of heuristics to determine the optimal value of 
$n$; ACL2(ml) has its own function to dynamically adjust $n$ for every run of clustering using an auxiliary \emph{granularity}  parameter. Granularity can be varied by the user;
and ranges between $1$ and $5$, low granularity produces big and general clusters while high granularity produces small and precise clusters (see the top row of Table~\ref{tab:compare}).
Given a granularity value $g$, the number of clusters $n$ is given by the formula {\small $$n=\lfloor\frac{\text{objects to cluster}}{10-g}\rfloor.$$}
The clustering algorithm assigns a \emph{proximity value} to every term in a cluster.
This ranges from $0$ to $1$, and indicates the certainty of the given example belonging to the cluster. 
ACL2(ml) shares with ML4PG some additional heuristics to ensure output quality, e.g., all experiments are run 200 times and only the ones with high frequencies are displayed to the user. 


The choice of clustering algorithm, granularity and ACL2 libraries are accommodated in ACL2(ml) through a menu included in the Emacs interface. 
On user's demand, the ACL2(ml) interface displays families of related 
theorems; we call this process \emph{Theorem Clustering}. 
In addition, ACL2(ml) clusters all the library definitions in the background every time a new definition is introduced. We call this process \emph{Definition Clustering}. 

The main reason for distinguishing theorem clustering and definition clustering is as follows. Theorem clustering is the ultimate goal
of the proof mining here, but the feature tables for the theorems depend on the numeric representation of the symbols appearing in the theorems (see Definition~\ref{df:matrix} and Table~\ref{tab:tt}).
These symbols are normally defined within the libraries one uses, or else imported from CLISP. The following definition proceeds inductively on the type of symbols appearing in ACL2 definitions.


\begin{figure}[t]
\centering
\begin{lstlisting}[frame=lines,mathescape,basicstyle=\tiny,breaklines=true]  
$\ast$ Type recognisers ($r=\{$symbolp, characterp, stringp, consp, acl2-numberp, integerp, rationalp, complex-rationalp$\}$): $[r_i]=1+\sum_{j=1}^i \frac{1}{10\times 2^{j-1}}$ (where $r_i$ is the $i$th element of $r$).
$\ast$ Constructors ($c=\{$cons, complex$\}$): $[c_i]=2+\sum_{j=1}^i \frac{1}{10\times 2^{j-1}}$.
$\ast$ Accessors ($a^1=\{$car, cdr$\}$, $a^2=\{$denominator, numerator$\}$, $a^3=\{$realpart, imagpart$\}$): $[a^j_i]= 3 + \frac{1}{10\times j} + \frac{i-1}{100}$.
$\ast$ Operations on numbers ($o=\{$unary-/, unary--, binary-+, binary-*$\}$): $[o_i]=4+\sum_{j=1}^i \frac{1}{10\times 2^{j-1}}$.
$\ast$ Integers and rational numbers: $[0]=4.3$, $[n] =4.3 + \frac{|n|}{10}$  (with $n\neq 0$ and $|n|<1$) and $[n]=4.3 + \frac{1}{100*|n|}$ (with $n\neq 0$ and $|n|\geq 1$).
$\ast$ Boolean operations ($b=\{$equal, if, <$\}$):  $[b_i]=5+\sum_{j=1}^i \frac{1}{10\times 2^{j-1}}$.
\end{lstlisting}
\caption{{\scriptsize \textbf{Formulas to compute the value of function $[.]$ for the ACL2 functions imported from CLISP.}
The above formulas serve to assign closer values to the functions within each of the six above groups, and more distant numbers 
across the groups -- thus distinguishing the groups unambiguously.}}\label{tab:builtins}
\end{figure}

\begin{definition}[{Function $[.]$}]\label{def:recurrent}
Given the $n$th term definition of the library (call the term $t$), a function $[.]$ is inductively defined for every symbol $s$ in $t$ as follows:\\
$-$ $[s] = i$, if $s$ is the $i$th distinct variable in $t$ (note that all formulas are implicitly universally quantified in ACL2);\\
$-$ $[s] = -[m]$, if $t$ is a recursive definition defining the function $s$ with measure function $m$ automatically assigned by the ACL2;\\ 
$-$ $[s]= k$ , if $s$ is a function imported from CLISP; and $[s]= k$ in Figure~\ref{tab:builtins};\\
$-$ $[s] = 5 + 2 \times j + p $, where $C_j$ is  a cluster obtained as a result of definition clustering with granularity $3$ for library definitions $1$ to $n-1$, 
$s\in C_j$ and $p$ is the proximity value of $s$ in $C_j$.
(Note that a cluster in definition clustering is given by a set of terms;  and the default granularity $3$ generally provides 
a good balance between the size of clusters and their precision.)

\end{definition}

Note the recurrent nature of clustering in Definition~\ref{def:recurrent}, with symbol numbering for the $n$th term depending on the clustering results for previous $n - 1$ terms.
As the above definition implies, the function $[.]$ is adaptive, 
and is recomputed automatically when new definitions (and hence new symbols) are 
introduced. The motivation behind the various parameters of Definition~\ref{def:recurrent} is as follows:\\
$-$ \emph{Variables.} 
The variable encoding reflects the number and order of unique variables appearing in the term, note its correspondence  to the De Bruijn 
indexes.\\
$-$  \emph{Recursive case.} For every recursive function $s$, ACL2 assigns a termination measure function $m$. So, $m$ necessarily exists for all recursive definitions 
and implicitly contains some ``type'' information (e.g. the measure for the function \verb"rev" is the length of its input and the measure for \verb"fact" is the value of 
its input). 
It has a negative value in order for feature values to distinguish the occurrence of the inductive symbol being currently defined from occurrence of any external functions invoked in the body of the term. The value $-[m]$ identifies all inductively defined symbols with the same ACL2 termination measure function.\\
$-$ 
Finally, the formula $5 + 2 \times j + p$ assigns $[s]$ a value within $[5+ 2 \times j, 5+ 2 \times j + 1]$, depending on their statistical proximity $p$ for that cluster -- $p$ always lies within $[0,1]$. Thus, 
elements of the same cluster have closer values comparing to  the values assigned to elements of 
other clusters or to the imported CLISP functions.

We finish this section with some examples of ACL2(ml) clustering. All examples are run with several clustering functions and a choice of statistical parameters, allowing us to evaluate whether the feature extraction method we present here is robust across a range of algorithms, see Table~\ref{tab:compare}. 

%
%
%
%
%
%
%
%

\begin{example}\label{ex:example5}
Using the \emph{JVM} library, ACL2(ml)  detects lemmas similar to theorem \verb"fib-fib-tail" (see Figure~\ref{fig:rec-tail}).
Table~\ref{tab:compare} shows the sizes of clusters that ACL2(ml) will display for this theorem for various clustering algorithms and granularities. In Table~\ref{tab:compare}, we can see a clear pattern: \verb"fib-fib-tail" is consistently grouped with other theorems related 
to the equivalence of recursive and tail-recursive functions -- this is done with all variations of learning algorithms and granularities, 
albeit with varied degree of precision.
For this set of examples and this stage of the proof, the feature extraction function $[.]$ returned values:\\
 $[\verb"fact"]= 12.974$, $[\verb"power"]= 12.973$, $[\verb"fib"] = 12.618$.\\
 $[\verb"helper-fact"] = 16.961$, $[\verb"helper-power"]= 16.967$, $[\verb"helper-fib"] = 16.431$.\\
$[\verb"fact-tail"]= 18.970$, $[\verb"power-tail"] = 18.969$, $[\verb"fib-tail"] = 18.735$.\\
Note that the numbers above correspond to our intuitive grouping in Figure~\ref{fig:rec-tail}.

Table~\ref{tab:compare} also shows the lemmas of library \emph{Lists} that are similar to \verb"revrev". 
In this case, the most precise cluster is detected by K-means with $g=5$, but we can notice that \verb"revrev" is always clustered with theorems \verb"sortsort" 
and \verb"int-x-x".

\begin{table}[t]
\centering
\scriptsize{
  \begin{tabular}{c|c|c|c|c|c|c|}
   \hhline{~------}
&	\multirow{2}{*}{Algorithm:}     & $g=1$ & $g=2$&$g=3$& $g=4$&$g=5$\\
&      & ($n=16$) & ($n=18$)&($n=21$)& ($n=25$)& ($n=30$) \\
\hline
\multirow{2}{*}{fib-fib-tail experiments} &   K-means  &$9^{a,b,c}$ & $4^{a,b,c}$  &$3^{a,c}$ &$2^{a}$ &  $2^{a}$ \\
   \hhline{~------}
\multirow{2}{*}{(one relevant cluster)}  &   E.M. &  $16^{a,b,c}$ &  $16^{a,b,c}$&  $9^{a,b,c}$ & $4^{a,b,c}$ &  $4^{a,b,c}$\\
   \hhline{~------}
  &   FarthestFirst & $12^{a,b,c}$ &$12^{a,b,c}$  & $10^{a,b,c}$  & $5^{a,b,c}$  & $4^{a,b,c}$  \\
\hline 
     \end{tabular} 
     
  \begin{tabular}{c|c|c|c|c|c|c|}
\hhline{~------}
&	\multirow{2}{*}{Algorithm:}     & $g=1$ & $g=2$&$g=3$& $g=4$&$g=5$\\
&      & ($n=11$) & ($n=12$)&($n=14$)& ($n=16$)& ($n=20$) \\
\hline
\multirow{2}{*}{~revrev experiments~}   &   K-means  & $22^{d,e}$  & $22^{d,e}$  & $20^{d,e}$ & $11^{d,e}$ & $3^{d,e}$  \\
   \hhline{~------}
 \multirow{2}{*}{~(one relevant cluster)~}  & E.M. & $29^{d,e}$  & $25^{d,e}$ &  $25^{d,e}$ & $23^{d,e}$ &  $19^{d,e}$ \\
   \hhline{~------}
 & FarthestFirst & $45^{d,e}$  & $34^{d,e}$   & $30^{d,e}$  & $26^{d,e}$ & $15^{d,e}$  \\
   \hline
     \end{tabular}} 
  
  \caption{\scriptsize{\emph{Results of Clustering experiments in ACL2(ml)} for a choice of algorithms and granularities. When granularity $g$ is chosen by the user,
  ACL2(ml) dynamically calculates the number $n$ of clusters; the table shows the size of one most relevant cluster that ACL2(ml) displays to the user in each case.
  \textbf{Top table}. Experiments for Theorem \texttt{fib-fib-tail} using JVM library (150 lemmas). We mark if the cluster contains (in addition to Theorem  \texttt{fib-fib-tail}):
  $a)$ Theorem  \texttt{fact-fact-tail}, $b)$ Theorem \texttt{power-power-tail},
  $c)$ other theorems related to the equivalence of recursive and tail-recursive functions. 
  \textbf{Bottom table}. Experiments for Theorem \texttt{revrev} using Lists library (100 lemmas). All the clusters contain theorems $d)$ \texttt{sortsort} and $e)$ \texttt{int-x-x}
  (in addition to \texttt{revrev}). Note the stable performance of our feature extraction algorithm across several algorithms and parameters.}}\label{tab:compare}
	\vspace{.5cm}
 \end{table}

\end{example}

\section{Lemma Discovery in ACL2(ml)}\label{sec:lemma-analogy}


One of the motivations of this research is to provide the ACL2 users with an efficient interactive lemma
suggestion mechanism when automated proof search fails. The symbolic side of ACL2(ml) uses analogical reasoning to efficiently produce lemmas which are relevant to the current conjecture. In particular, it automatically attempts to suggest lemmas which 
are characterised as \emph{eureka lemmas} -- i.e. ones whose invention is
mathematically creative and difficult to automate. 
In Section~\ref{sec:acl2ml}, we described how ACL2(ml) employs statistical machine-learning
to determine similarity of theorem statements based on the term structure. 
In this section, we show how ACL2(ml) uses this information to
construct analogous lemmas -- this greatly reduces the search space for lemma discovery.

\begin{figure}[!t]
\begin{minipage}{0.5\textwidth}
\centering
\includegraphics[width=0.98\textwidth]{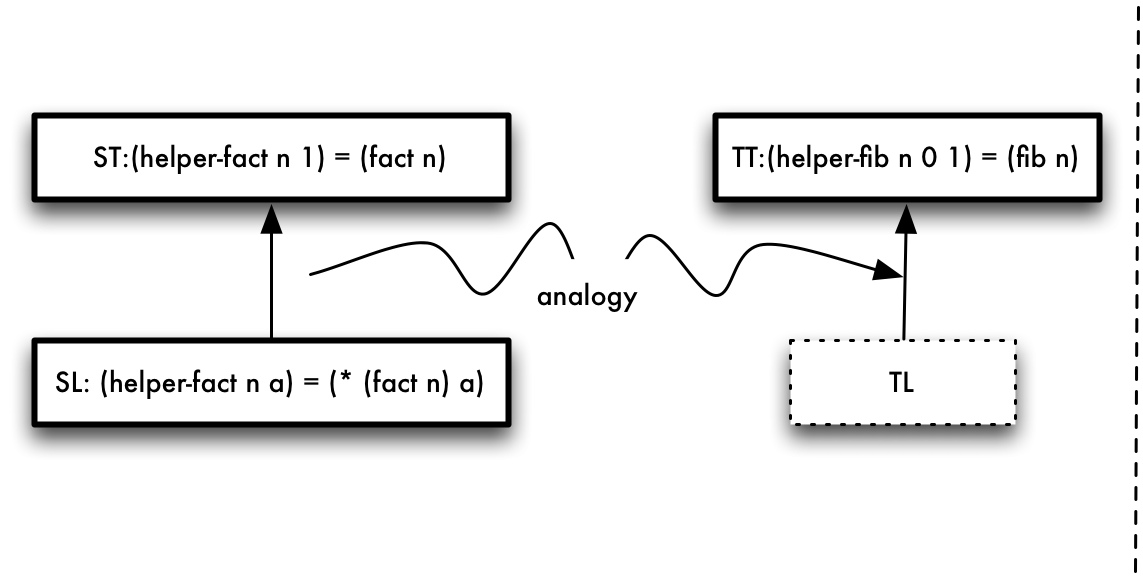}
\end{minipage}
\begin{minipage}{0.5\textwidth}
\centering
\includegraphics[width=0.98\textwidth]{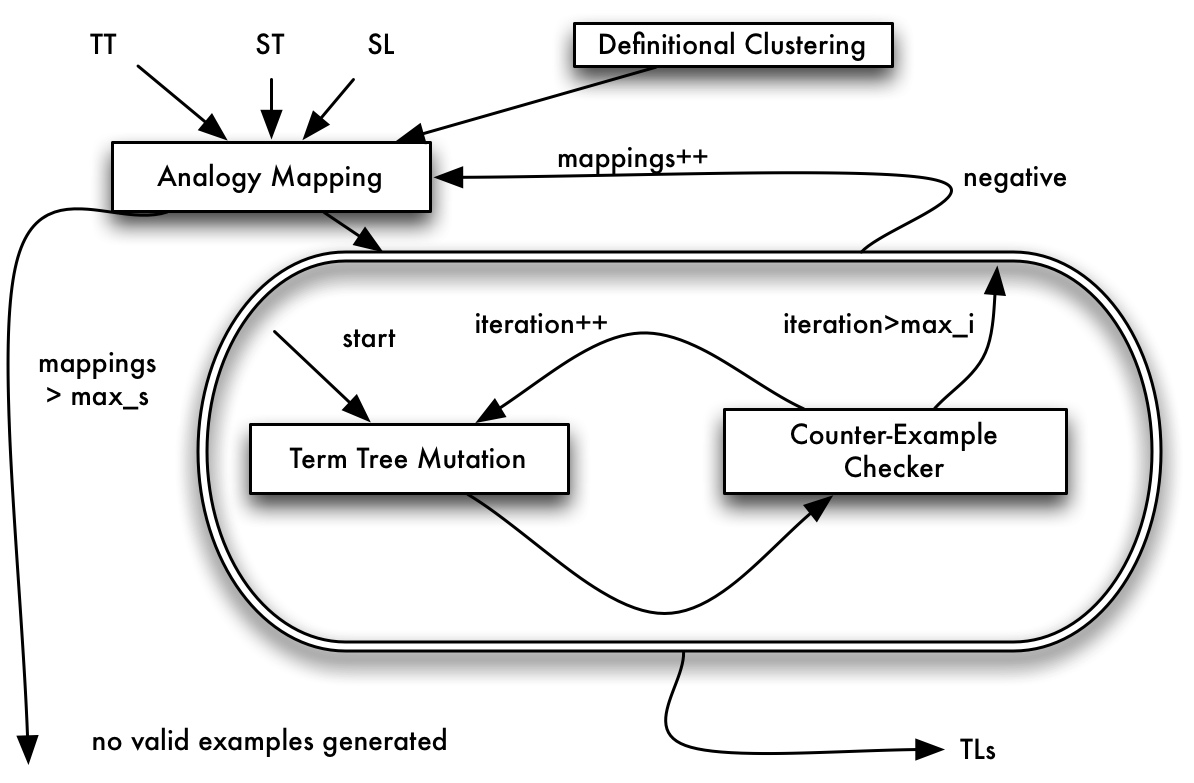}
%
\end{minipage}
\caption{\scriptsize{\emph{
    Left: The source theorem and lemma along with the target theorem for which we seek an analogous lemma. Right: overview of the process of generating analogical lemmas. Here the label  {\tt iteration} is the iteration step  
    and {\tt mappings} is a count of the number of analogy mappings.}}
\label{fig:slcom}}
\end{figure}

We introduce the following terminology:
A \emph{Target Theorem (TT)} is a theorem currently being attempted in
  ACL2, but requiring user's intervention. 
A \emph{Source Theorem (ST)} is a theorem which has been suggested as similar to the TT
  by the statistical ACL2(ml).
A \emph{Source Lemma (SL)} is a user-supplied lemma required for proving the source theorem.
The symbolic side of ACL2(ml) groups (potentially multiple) statistical suggestions into 
 ST and SL pairs, each being
evaluated in turn. The process then outputs some \emph{Target Lemmas (TLs)} --- these lemmas are analogical to some SL and not falsified by 
counter-example checking. 
\begin{example}\label{ex:runex}
Let us consider the case shown by Figure \ref{fig:slcom}, 
where a user of ACL2 wants to prove the TT\footnote{To ease readability, we use infix notation instead of ACL2 notation, e.g. \texttt{t1 = t2} instead of \texttt{(equal t1 t2)}.}: 
{\small \verb"(helper-fib n 0 1) = (fib n)"}. As described in Section \ref{sec:acl2ml} and Table \ref{tab:compare}, ACL2(ml) suggests the closest analogy to be:\\ {\small \verb"(helper-fact n 1) = (fact n)"}.
Only one user-defined auxiliary lemma is used in the proof of this ST: 
{\small \verb"(helper-fact n a) = (* a (fact n))"}. The job of the lemma analogy process is to
construct the corresponding TL for the \texttt{fibonacci} example, this will be our running example throughout this section. 
\end{example}

The overall process for lemma analogy is shown in Figure
\ref{fig:slcom}. It has two main components: \emph{Analogy Mapping} and \emph{Term Tree Mutation}. Analogy Mapping calculates which symbols could be analogical to each other using the definition clustering techniques from Section~\ref{sec:acl2ml}. Term Tree Mutation then uses this information to construct candidate target lemmas analogical to a given SL.  
Symbols belonging to the background theory, i.e. symbols shared between the source and target, are not changed by Analogy Mapping.
\begin{definition}[Analogy Mapping $\cal A$] \label{def:analogy-map}
For all symbols  $s_1, \ldots, s_n $ occurring in the current ST, the set 
of admissible analogy mappings is the set of all mappings ${\cal A}$ 
such that\\ 
- ${\cal A}(s_i) = s_i$ for all shared background symbols; otherwise:\\
- ${\cal A}(s_i) = s_j$ for all combinations of $i,j \in 1 \ldots n$, such that $s_i$ and $s_j$ belong to the same cluster in the last iteration of definition clustering.
\end{definition}

\begin{example}\label{ex:mapping}
For our running example, the shared background theory includes symbols \{\texttt{+}, \texttt{*},\texttt{-}, \texttt{1}, \texttt{0}\}. We thus get a mapping:  \\
${\cal A}$ = \{{\tt fact $\mapsto$ fib}, {\tt helper-fact $\mapsto$ helper-fib}, {\tt + $\mapsto$ +}, {\tt 1 $\mapsto$ 1} $\ldots$ \}
\end{example}




\textbf{Term Tree Mutation.} 
Given an analogy mapping, candidate lemmas are constructed by mutating the term tree of the SL.
IsaScheme \cite{montano2012} works similarly, by defining {\em schemes} which determine the shape of a generated term. Here, we extend this idea by iteratively allowing larger perturbations on this shape if no suitable candidate lemma has been found. We refer to this as {\em mutation}, an overview of the process is shown in Figure \ref{fig:slcom}. 
The algorithm proceeds down three levels of increasing term tree mutation. After each iteration, we test the
validity of the set of generated equations using a counter-example checker. If no candidate conjecture survives 
counter-example checking, mutation proceeds to the next iteration. The Term Tree Mutation algorithm for the case of equational lemmas is shown in Figure \ref{fig:TTMut}, however note that each sub-routine work on arbitrary terms and the mutation algorithm can thus trivially be generalised. The first iteration, \emph{Tree Reconstruction}, replaces symbols in the SL
with their analogical counterparts. The second iteration, \emph{Node Expansion} 
further mutates the term, by synthesising small terms (max depth 2) in place of variables. Finally, the last iteration, \emph{Term Tree Expansion} 
similarly adds new term structure, but on the top-level of the term.

\begin{figure}[t]
\centering
\begin{Verbatim}[frame=single,fontsize=\scriptsize,commandchars=\\\{\}]  
TTMutation
Input:  An analogy mapping A, a set of shared symbols F, and an equational source 
        term tl = tr. 

\textit{\ 1. Compute Tree Reconstructions giving sets of candidate left- and right hand sides.}		  
L1 = TreeRec(A, tl)
R1 = TreeRec(A, tr)
		  
\textit{\ Counter-example check all pairs of terms as candidate equations.}
Res = TestAllEqs(L1, R1)
if not(Res = []) then return Res 
\textit{\ If no candidate lemma found, continue to next iteration.}
else	 

\textit{\ 2. Compute Node Expansions}
L2 = map (NodeExp F) L1
R2 = map (NodeExp F) L2

\textit{\ Counter-example check all pairs of terms generated so far as candidate equations.}
Res = TestAllEqs(L1++L2, R1++R2)
. . .
	
\textit{\ 3. Compute Tree Expansions}
L3 = (TreeExp F) L2
R3 = (TreeExp F) R2		  

return TestAllEqs (L1++L2++L3, R1++R2++R3)
\end{Verbatim}
\vspace{-.5cm}
 \caption{{\scriptsize Term Tree Mutation. The algorithm proceeds through three iterations of increasingly mutating the source term. This work is performed by the functions \texttt{TreeRec}, \texttt{NodeExp} and \texttt{TreeExp}. The left- and right hand sides of the original equation are mutated separately and after each step all combinations of equalities are counter-example tested (by the function \texttt{TestAllEqs}). }}
 \label{fig:TTMut}
\end{figure}

\textbf{Iteration 1 -- Tree Reconstruction.} 
The first iteration recursively replaces symbols in the source term by their analogical counterpart specified by $\cal A$.
If the analogous function has fewer arguments than the source, we simply ignore excess arguments. If it has more arguments, we insert variables in the free positions.

\begin{example}
The running example has SL {\small \verb"(helper-fact n a) = (* a (fact n))"} with 
$\cal A$ as in Example~\ref{ex:mapping}. \texttt{TreeRec} performs 
the replacements to produce sets of candidate left- and right-hand sides, including for instance, the potential left-hand side  {\small \verb"(helper-fib n n1 a)"} and the right-hand sides {\small \verb"(* a (fib n))"} and {\small \verb"(* n1 (fib a))"}
-- as well as other variants with variables in different order. Note that we introduce a new variable 
\texttt{n1} as {\small \verb"helper-fib"} has one more argument than its analogical counterpart {\small \verb"helper-fact"}.
In this case, no combination of left- and right-hand sides yield an equational theorem. 
\end{example}

\textbf{Iteration 2 -- Node Expansion.} 
If Tree Reconstruction fails to produce any conjecture passing counter-example testing, the second iteration synthesises
terms which are allowed to replace variables, thus growing the term tree from the leaves. These synthesised terms are however limited to 
depth 2 and built from the shared function symbols, to keep the search space tractable.

\begin{example}
In our running example, we look at the new terms generated from the right-hand side of the SL.
Suppose we start from one candidate from the previous iteration: {\small \verb"(* n1 (fib a))"}.
Node Expansion will consider all possible ways of replacing the variables \texttt{a} and/or \texttt{n1} 
by terms built from shared background theory functions \{\texttt{+}, \texttt{*}, \texttt{-}\}, applied to available variables, \{\texttt{a}, \texttt{n}, \texttt{n1}\}, 
and constants, \{\texttt{1}, \texttt{0}\}. The set of potential right-hand sides will now include terms such as 
{\small \verb"(* n1 (fib (- n 1)))"} (having replaced \texttt{a} by \texttt{(- n 1)}), as well as many more similar alternatives.  
\end{example}

\textbf{Iteration 3 -- Term Tree Expansion} 
As opposed to Iteration 2, which introduced new term structure at the leaves, this phase allows for insertion of term
structure at the top level. Term Tree Expansion considers all terms synthesised so far and explores how they can be used 
as arguments to shared background theory functions.

\begin{example}
For the running example, we consider the whole set of candidate terms for the right-hand side. We consider all terms
that can be built giving some combination of these as arguments to the background theory functions \{\texttt{+}, \texttt{*}, \texttt{-}\}. 
For instance, one of the new terms that is built in this iteration is {\small \verb"(+ (* n (fib (- n1 1))) (* a (fib n1))"}, 
having added a top-level \texttt{+}. The counter-example checker does not find a counter-example 
for the conjecture
{\small \verb"(helper-fib n n1 a) = (+ (* n (fib (- n1 1))) (* a (fib n1))"}. Hence, 
the result is shown to the user as a suggestion of an auxiliary lemma to employ in the current
proof.
\end{example}

\section{Evaluation and Conclusions}\label{sec:concl}

We have presented ACL2(ml), 
an ACL2 extension combining statistical machine learning, to detect proof patterns, with symbolic techniques for generating new auxiliary lemmas. All software and extensive evaluation results are available from~\cite{acl2ml}.
Both sides of ACL2(ml) are original techniques on their own. However, comparing ACL2(ml) with other alternative tools, two of its features distinguish it from all other approaches: 
its methods of generating the proof-hints \emph{interactively} and in real-time, in response to the user's call; 
and secondly, its flexible environment for integration of statistical and symbolic techniques.
We finish with an evaluation of the system. 

\textbf{Scalability.}
The ACL2 proofs  libraries can grow very big, especially in industrial scenarios. The statistical ACL2(ml) tool works well with libraries of varied sizes and complexities; and does not need any fine-tuning when the user adds more libraries. 
To illustrate this, we increased by  $6.64$ times the data set used in Example~\ref{ex:example5} and Table~\ref{tab:compare} (by adding as noise lemmas coming from 
standard ACL2 libraries; and also the ones presented in \cite{Mateos09}
-- the data set contains 996 theorems). Table~\ref{tab:compare-big} shows that ACL2(ml) still finds precise clusters; e.g., 
it still finds a cluster containing exactly the three  theorems proving equivalence of recursive and
tail-recursive functions similar to  \verb"fib-fib-tail".  Similar results are shown with respect to other functions from e.g. JVM library, see~\cite{acl2ml}.

\vspace{-.5cm}
\begin{table}[!]
\centering

\begin{tabular}{c|c|c|c|c|c|}
   \hhline{~-----}
&	   $g=1$ & $g=2$&$g=3$& $g=4$&$g=5$\\
&       ($n=110$) & ($n=124$)&($n=142$)& ($n=166$)& ($n=199$) \\
\hline
fib-fib-tail  & $57^{a,b,c}$ &  $50^{a,b,c}$ &$25^{a,b,c}$ &$8^{a,b,c}$ &  $4^{a,b,c}$ \\
\hline
\end{tabular} 
  
  \caption{\scriptsize{\emph{Results of Clustering experiments in ACL2(ml)} for $k-$means clustering algorithm and a choice of granularities. Experiments for Theorem \texttt{fib-fib-tail}. 
  We mark if the cluster contains (in addition to Theorem \texttt{fib-fib-tail}): 
  $a)$ Theorem \texttt{fact-fact-tail}, $b)$ Theorem \texttt{power-power-tail}, $c)$ theorems related to the equivalence of recursive and tail-recursive functions. 
  The size of the data set is $996$ theorems coming from the JVM library~\cite{Moore03provingtheorems}, the standard ACL2 libraries:
Lists, powerlists and sorting; and a library related to the formalisation of a computer algebra system presented in~\cite{Mateos09}.}}\label{tab:compare-big}
 \end{table}
 
\vspace{-.75cm} 
\textbf{Usability of statistical suggestions.} The statistical side of ACL2(ml) is not deterministic; but  it is possible to empirically evaluate the usability of the statistical results
produced by ACL2(ml). Using the data set containing 996 theorems presented in Table~\ref{tab:compare-big}, and considering the clusters obtained with 
granularity $5$, we notice that: $37\%$ of clusters identified by ACL2(ml) can be directly used by the Lemma Analogy tool of ACL2(ml) to mutate lemmas. Additionally, $15\%$ of clusters contain theorems that use the same lemmas in their proofs, another $15\%$ of clusters consist of theorems that are used in the proofs of other theorems of the same cluster, and about $19\%$ of clusters contain basic theorems whose proofs are similar, but based on simplification, and hence unsuitable for Lemma Analogy application. 
Only $14\%$ of clusters do not show a clear correlation that could be reused.

\textbf{Modularity.} ACL2(ml) provides a flexible environment for integrating statistical and symbolic machine-learning methods.
The statistical parameters can be easily tuned to provide a better input for the Analogy tool.

\begin{example}\label{ex:example6}
 Consider the generation of auxiliary lemmas for \verb"fib-fib-tail" and \verb"revrev". The auxiliary lemma for \verb"fib-fib-tail" 
 is generated from the lemma marked with either $a$ or $b$ in Table~\ref{tab:compare}. 
 Other lemmas in this cluster can be considered as noise. In this case, the granularity and algorithm 
 choice are not particularly relevant because the auxiliary lemma can be created from all the clusters and the search space is small. 
 However, the correct auxiliary lemma is generated faster when increasing the granularity parameter. 
 
 In the case of \verb"revrev", the relevant clusters are bigger than in the case of \verb"fib-fib-tail"
 and the relevant lemma for \verb"revrev" is only generated from theorem \verb"sortsort". Then, the noise can be reduced by setting ACL2(ml) to the highest granularity $5$ with K-means algorithm, which reduces the cluster to $3$ lemmas. 
  \end{example}

All our running examples were based on the cases when the proofs of the theorems required introduction of only one user-defined auxiliary lemma; and hence the analogy tool concentrated only on them. However, in the general case, the similar theorems found by statistical ACL2(ml) can rely on several auxiliary lemmas. 
To help the Analogy tool decide which of them should be used for new lemma generation, the statistical tool can be applied recursively to cluster these auxiliary lemmas; and thus find those more likely to result in a good analogy.

\textbf{Lemma Discovery.}   The Analogy algorithm is designed to be fast and reduce the combinatorial explosion
introduced by typical theory exploration techniques. By starting from
an exemplar source lemma, we use mutation to perturb the term.
We compare our tool with QuickSpec \cite{quickspec} -- a state of the art system for generating equational
conjectures about Haskell programs. While QuickSpec is designed to suggest a number of
conjectures about a given set of  symbols, the lemma analogy system
aims at just discovering a few lemmas relevant to some
particular proof attempt. 

The ACL2(ml) approach is more suitable for integration into an ITP, as analogy can reduce the search space, making lemma discovery tractable in theories where QuickSpec runs out of memory. QuickSpec works by generating all terms up to a given depth (default is 3) and then use testing to evaluate the terms and divide them into equivalence classes. These represent a large number of equations so the set is pruned before displaying it to the user in an attempt to only show a small set of interesting equations. 
Table \ref{tab:qs} shows the results of all combinations of source and target theorems from the examples
used on the natural numbers examples introduced into ACL2(ml). QuickSpec takes as input only the function symbols from the target theorems for each run. Both systems produce results instantly, except in the cases where QuickSpec runs into exponential blowups.
The Analogy tool usually returns one lemma relevant to the TT,
whereas QuickSpec may return a very large number of results (some of which are not valid), which is not ideal for displaying to a user of an interactive system. It also runs out of memory on examples involving functions with many arguments  and a higher degree of commutativity, such as the tail recursive version of multiplication (as the number of terms within its depth limit increases). These kind of theories are generally problematic for any theory exploration system relying merely on term generation, e.g. \cite{JDB11,montano2012,hipspec}. Despite their heuristics, the search space grows too big. The integration in an ITP allows us to use information about the current proof attempt, and thus navigate these theories.  

Like proof-critics \cite{BB05}, which is a technique for lemma discovery in automated provers, we use information from the current proof-attempt to guide the 
search for auxiliary lemmas. However, while proof-critics only analyses the current proof attempt, ACL2(ml) also learns from all previous proofs. Some proof-critics
are also closely reliant on particular proof-techniques, such as rippling, while ACL2(ml) is independent of this. 

\begin{table}[t]
\centering
\begin{tabular}{cc|c|c|c|c|c|c|c|c|c|c|c|}
\cline{3-9}\cline{11-13}
& & \multicolumn{7}{ c| }{Target} &\mbox{\hspace{0.2cm}}&\multicolumn{3}{c|}{QuickSpec} \\ \cline{3-9}\cline{11-13}
& &$fact$&$power$&$expt$&$sum$& $sum\_sq$&$mult$&$fib$&&Lemma&Valid&Invalid  \\ \cline{1-9}\cline{11-13}
\multicolumn{1}{|c|}{\multirow{7}{*}{Source}} &
\multicolumn{1}{|c|}{$fact$}&-&\checkmark$_1$&\checkmark$_1$&\checkmark$_1$&\checkmark$_1$&\checkmark$_2$&\checkmark$_1$&&$\times$&10&5 \\ \cline{2-9} \cline{11-13}
\multicolumn{1}{|c}{}&
\multicolumn{1}{|c|}{$power$}&\checkmark$_1$&-&\checkmark$_1$&\checkmark$_1$&\checkmark$_1$&\checkmark$_2$&\checkmark$_1$&&$\times$&17&4 \\ \cline{2-9} \cline{11-13}
\multicolumn{1}{|c}{}&
\multicolumn{1}{|c|}{$expt$}&\checkmark$_1$&\checkmark$_1$&-&\checkmark$_1$&\checkmark$_1$&\checkmark$_2$&\checkmark$_1$&&$\times$&OoM&OoM \\ \cline{2-9} \cline{11-13}
\multicolumn{1}{|c}{}&
\multicolumn{1}{|c|}{$sum$}&\checkmark$_1$&\checkmark$_1$&\checkmark$_1$&-&\checkmark$_1$&\checkmark$_2$&\checkmark$_1$&&$\times$&7&2\\ \cline{2-9} \cline{11-13}
\multicolumn{1}{|c}{}&
\multicolumn{1}{|c|}{$sum\_sq$}&\checkmark$_1$&\checkmark$_1$&\checkmark$_1$&\checkmark$_1$&-&\checkmark$_2$&\checkmark$_1$&&$\times$&7&1 \\ \cline{2-9} \cline{11-13}
\multicolumn{1}{|c}{}&
\multicolumn{1}{|c|}{$mult$}&\checkmark$_1$&\checkmark$_1$&\checkmark$_1$&\checkmark$_1$&\checkmark$_1$&-&\checkmark$_1$&&$\times$&~200&~20\\ \cline{2-9} \cline{11-13}
\multicolumn{1}{|c}{}&
\multicolumn{1}{|c|}{$fib$}&$(\checkmark_2)$&$(\checkmark_2)$&$\times$&$\times$&$\times$&$\times$&-&&$\times$&OoM&OoM\\ \cline{1-9}\cline{11-13}
\end{tabular}
\caption{{\scriptsize{\emph{Results of lemma analogy in comparison with
      QuickSpec. \checkmark$_n$ denotes a success with $n$ returned
      lemmas, and (\checkmark$_n$) denotes a valid generated lemma
      which is not applicable to the TT. $\times$ denote failure to find exactly the desired lemma, while OoM denotes that the system
      ran out of memory. The numbers correspond to the number of
      formulae generated by QuickSpec}}}}
\label{tab:qs}
\end{table}

\textbf{Limitations and Further Work.}
Below, we summarise the limitations of the system and indicate directions of further work.
 Although the statistical side of ACL2(ml) can process and cluster any proof and lemma shape, 
the symbolic analogy tool cannot handle all possible analogical cases (and no algorithmic approach could do that in principle).

\emph{Different patterns}. 
 Statistical ACL2(ml) groups in the same clusters theorems \verb"revrev", and \verb"int-x-x"; however, the lemmas used in 
 \verb"revrev" cannot be mutated to generate any of the lemmas needed in the proof of \verb"int-x-x".

\emph{Smaller lemmas}. The lemma analogy tool currently only adds term structure; therefore, it cannot generate smaller lemmas. E.g. 
  Lemmas \verb"fact-fact-tail" and \verb"fib-fib-tail" are in the same cluster (see Table~\ref{tab:compare}); the analogy tool succeeds in generating the lemma \verb"fib-fib-tail" given the source lemma \verb"fact-fact-tail", but fails to go the other way around. This would require removing term structure, i.e. the inverse operation of Term Tree Expansion (see Section~\ref{sec:lemma-analogy}). 
  
  

\emph{Conditional lemmas}. Apart from type recogniser conditions (e.g. \verb"natp" is the type recogniser for natural numbers), 
  several ACL2 lemmas need additional conditions, which, at the moment, are not generated by the symbolic ACL2(ml). Discovering appropriate conditions for generated lemmas is a difficult problem for theory exploration systems.   

\emph{New definitions}. Another big challenge in lemma discovery is invention of new concepts. 
 The proof of Lemma \verb"int-x-x" needs a new concept called \verb"subp" which is the recogniser for subsets of lists -- \verb"(subp x y)" returns \verb"true" or \verb"false" according to whether every element of \verb"x" is an element of \verb"y".



\bibliographystyle{plain}
\bibliography{lpar13}

\appendix

\section{ACL2(ml) interface}

\begin{figure}
 \centering
 \includegraphics[scale=.31]{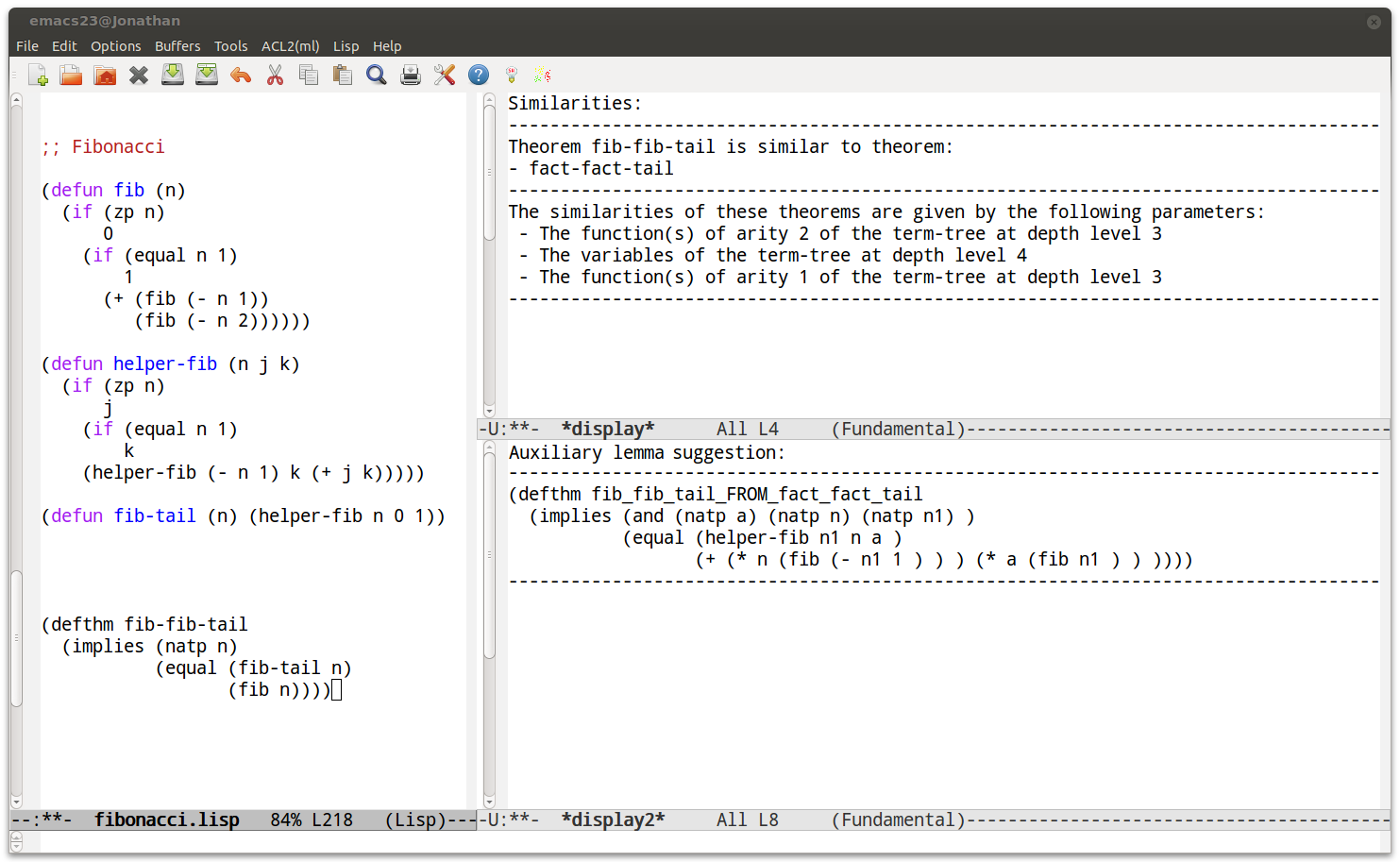}
 \caption{{\scriptsize {\emph ACL2(ml) interface.} The Emacs window has been split into three regions: the left one keeps the 
 ACL2 script, the top-right one shows the output of statistical ACL2(ml): theorems which are similar to \texttt{fib-fib-tail} (see Section~\ref{sec:acl2ml}) and also the  correlation of which features formed the pattern.  The bottom-right window shows the output of the symbolic ACL2(ml): suggestion
 to add and prove auxiliary lemma \texttt{fib-fib-tail} (see Section~\ref{sec:lemma-analogy}).}}\label{fig:screenshot}
\end{figure}

ACL2(ml) has been developed as an extension for Emacs, see Figure~\ref{fig:screenshot}. This extension includes a new menu
called \emph{ACL2(ml)} that allows the user to configure the different parameters: algorithm, granularity and libraries to 
cluster. 

ACL2(ml) offers two functionalities which can be invoked from the two most right buttons of the interface. The right most button 
clusters libraries irrespective of the theorem that the user is proving. This functionality can be useful to obtain an overview of 
the similar lemmas which have been developed in the different libraries. The other button provides the functionality described in this paper: 
given a current theorem $T$, ACL2(ml) obtains the theorems that are similar to $T$ and invokes the lemma analogy tool to obtain auxiliary lemmas.

\begin{example}
 In Figure~\ref{fig:screenshot}, ACL2(ml) shows both the theorems which are similar to \verb"fib-fib-tail" (see Figure~\ref{fig:rec-tail}) and the suggestion provided for this theorem. 
 We use the following parameters:
 
 \begin{itemize}
  \item Data set: JVM library.
  \item Algorithm: K-means.
  \item Granularity: 5.
 \end{itemize}

\end{example}

\newpage
\section{Formalisation of JVM bytecode in ACL2}\label{ap:jvm}

In~\cite{Moore03provingtheorems}, an interpreter for JVM programs, called TJVM, was modelled in ACL2. Java 
bytecode can be executed within TJVM and the correctness of Java bytecode can be proven using ACL2. For instance,
in the case of the factorial program (see Figure~\ref{fig:factorial}) , we can prove the following theorem, which states the correctness of the
factorial bytecode. 

\begin{lemma}\label{lemma:jvm}
Given a natural number $n$ and the factorial program with $n$ as an input, TJVM produces a state which contains $n!$ on top of the stack running the
bytecode associated with the program.

\begin{figure}
\begin{minipage}{0.5\linewidth}
\centering
{\small
\begin{lstlisting}
static int factorial(int n)
{
  int a = 1;
  while (n != 0){
    a = a * n;
    n = n-1;
    }
  return a;
}
\end{lstlisting}}
\end{minipage}
\begin{minipage}{0.5\linewidth}
\centering
{\scriptsize
$\begin{array}{ccl}
0&:& iconst~1\\
1&:& istore~1\\
2&:& iload~0\\
3&:& ifeq~13\\
4&:& iload~1\\
5&:& iload~0\\
6&:& imul\\
7&:& istore~1\\
8&:& iload~0\\
9&:& iconst~1\\
10&:& isub\\
11&:& istore~0\\
12&:& goto~2\\
13&:& iload~1\\
14&:& ireturn\\
\end{array}$}
\end{minipage}
\caption{Factorial function. Left: Java program for computing the factorial of natural numbers. Right: Java bytecode associated with the Java program.}\label{fig:factorial}
\end{figure}
\end{lemma}

The proof of theorems like Lemma~\ref{lemma:jvm} always follows the same methodology which consists of the following three steps. 

\begin{enumerate}
 \item Write the specification of the function, write the algorithm, and prove that the algorithm satisfies the specification.
 \item Write the JVM program within ACL2, define the function that schedules the program (this function will make TJVM run the program to completion as
a function of the input to the program), and prove that the resulting code implements this algorithm.
 \item Prove total correctness of the Java bytecode. 
\end{enumerate}

The first step in the above methodology corresponds with the definitions and lemmas presented in Figure~\ref{fig:rec-tail} about the equivalence of 
recursive and tail recursive functions.

\begin{figure}
\centering
\begin{Verbatim}[frame=single,fontsize=\scriptsize,commandchars=\\\{\}]                
;; \textit{Factorial}
1 (defthm program-correct-fact
    (implies (natp n)
         (equal (run (sched-fact n) (make-state 0 (list n) nil  *pi-fact*))
                (make-state 14 (list 0 (fact-tail n)) (push (fact-tail n) nil) *pi-fact*))))

2 (defthm total-correctness-fact
    (implies (and (natp n)
                  (equal sf (run (sched-fact n) (make-state 0 (list n) nil *pi-fact*))))
             (and (equal (next-inst sf) '(HALT))
                  (equal (top (stack sf)) (fact n))))
                  
;; \textit{2^n}
1 (defthm program-correct-power
    (implies (natp n)
      (equal (run (sched-power n) (make-state 0 (list n) nil  *pi-power*))
             (make-state 14 (list 0 (power-tail n)) (push (power-tail n) nil) *pi-power*))))

2 (defthm total-correctness-power
    (implies (and (natp n)
                  (equal sf (run (sched-power n) (make-state 0 (list n) nil *pi-power*))))
             (and (equal (next-inst sf) '(HALT))
                  (equal (top (stack sf)) (power n))))
                  
;; \textit{Fibonacci}
1 (defthm program-correct-fib
    (implies (natp n)
          (equal (run (sched-fib n) (make-state 0 (list n) nil  *pi-fib*))
                 (make-state (if (equal n 0) 22 24) (fib-locals n 0 1) 
                                                    (push (fib-tail n) nil) *pi-fib*)

2 (defthm total-correctness-fib
    (implies (and (natp n)
                  (equal sf (run (sched-fib n) (make-state 0 (list n) nil *pi-fib*))))
             (and (equal (next-inst sf) '(HALT))
                  (equal (top (stack sf)) (fib n))))                  

\end{Verbatim}
\vspace{-.5cm}
 \caption{{\scriptsize \emph{Lemmas involved in the second and third steps of the methodology to prove the correctness of the JVM programs for factorial, $2^n$ and Fibonacci.} 
 1: theorems of Step 2 of the methodology. 2: theorems of Step 3 of the methodology. The complete 
 development can be downloaded from~\cite{Moore03provingtheorems}}}\label{fig:jvm}
\end{figure}

From this methodology, we can notice that there are 3 families of lemmas: one per each step of
the methodology (see Figures~\ref{fig:rec-tail} and~\ref{fig:jvm}). Later on, Table~\ref{tab:compare-big} shows that ACL2(ml) is able to find these three families of lemmas.

\newpage
\section{Term Tree Mutation Subroutines}
\label{ap:ttMut}

\begin{figure}[h]
\centering
\begin{Verbatim}[frame=single,fontsize=\tiny,commandchars=\\\{\}]  
TreeRec
Input: An analogy mapping A, a term t and a set of fresh variables V
	  
 case t of 
   Constant c: 
	return A(c)	
   Application fs(a1 \ldots an):
	\textit{Compute analogies for the function and its arguments}
	ft = A(fs)
	Args = map TreeRec [a1 \ldots an]

\textit{\ Source has equal or larger number of arguments, use (some of) the analogous arguments. }
if arity(fs) => arity(ft)	
	for all combinations of type correct arguments bi in Args:
	return set of terms ft(b1 \ldots bm) 	
\textit{\ Source has smaller number of arguments, pad with fresh variables. }	
if arity(fs) < arity(ft)
	for all combs of type correct arguments bi in Args or V:
	return set of terms ft(b1 \ldots bm) 		       
\end{Verbatim}

\begin{Verbatim}[frame=single,fontsize=\tiny,commandchars=\\\{\}]  
NodeExp
Input: A set of definitional symbols F, a term t and a set of variables V
	
case t of 
   Application fs(a1 \ldots an):
	fs(NodeExp a1 \ldots NodeExp an)

\textit{Any definitional constant may be replaced by another definitional constant of same type}	
   Constant c: 
   	if c is a member of F then
		return set of constant in F of same type as c.
	else return c 	
\textit{\ Expand a variable by inserting a function, applied to arguments drawn from F or V}   
   Variable v: 
   	for all symbols f in F of the same type as v:
		for all combinations of type-correct arguments bi in F or V:	
		return set of terms f(b1 \ldots bm)
\end{Verbatim}

\begin{Verbatim}[frame=single,fontsize=\tiny,commandchars=\\\{\}]  
TreeExp
Input: A set of terms T and a set of definitional symbols F

\textit{\ Apply each function to all type correct tuples of terms in T}	 
for each f in F: 
	build a set of terms Tf from all combinations of type correct ti's from T of form:
	f(t1 \ldots tn)
	
\textit{\ Return all possible function applications to terms from T.}	
return UNION (Tf\ldots Tg) 		 
\end{Verbatim}

 \caption{{\scriptsize \textbf{Top}. Term Tree Reconstruction. The algorithm replaces symbols in the given term, with their analogical counterparts.
 \textbf{Centre}. Node Expansion. The algorithm grows the term tree from the leaves, by possibly replacing constants by other constants, or variables by new depth 2 subterms.
 \textbf{Bottom}. Tree Expansion. The algorithm grows the term tree from the top, by picking a definitional function and adding previously generated terms as arguments.}}
  \label{fig:TTRec}
\end{figure}

\newpage
\section{Scaling ACL2(ml)}\label{ap:scaling}

From the methodology of Appendix~\ref{ap:jvm}, we can notice that there are 3 families of lemmas: one per each step of
the methodology (see Figures~\ref{fig:rec-tail} and~\ref{fig:jvm}). Table~\ref{tab:compare-big} shows that ACL2(ml) is able to find these three families of lemmas.
In this case, our data set for clustering consists of the following libraries: JVM library~\cite{Moore03provingtheorems}, the standard ACL2 libraries:
Lists, powerlists and sorting; and a library related to the formalisation of a computer algebra system presented in~\cite{Mateos09} (in total, the data set contains 996 lemmas).

\begin{table}[!]
\centering
  \begin{tabular}{c|c|c|c|c|c|}
   \hhline{~-----}
&	   $g=1$ & $g=2$&$g=3$& $g=4$&$g=5$\\
&       ($n=110$) & ($n=124$)&($n=142$)& ($n=166$)& ($n=199$) \\
\hline
fib-fib-tail  & $57^{a,b,c}$ &  $50^{a,b,c}$ &$25^{a,b,c}$ &$8^{a,b,c}$ &  $4^{a,b,c}$ \\
\hline
\hline
program-correct-fib &   $16^{d,e,f}$& $16^{d,e,f}$  &$15^{d,e,f}$ &$10^{d,f}$ & $3^{d,f}$   \\
   
\hline
   \hline
total-correctness-fib  &   $5^{g,h,i}$    & $5^{g,h,i}$  &$5^{g,h,i}$  &$5^{g,h}$ & $2^{h}$ \\
   \hline
     \end{tabular} 
  
  \caption{\scriptsize{\emph{Results of Clustering experiments in ACL2(ml)} for a choice of algorithms and granularities. 
  \textbf{Top table}. Experiments for Theorem \texttt{fib-fib-tail}. We mark if the cluster contains (in addition to Theorem \texttt{fib-fib-tail}): 
  $a)$ Theorem \texttt{fact-fact-tail}, $b)$ Theorem \texttt{power-power-tail}, $c)$ theorems related to the equivalence of recursive and tail-recursive functions. 
  \textbf{Centre table}. Experiments for Theorem \texttt{program-correct-fib}.  We mark if the cluster contains (in addition to Theorem \texttt{program-correct-fib}): 
  $d)$ Theorem \texttt{program-correct-fact}, $e)$ Theorem \texttt{program-correct-power}, $f)$ theorems related to Step 2 of the methodology.
  \textbf{Bottom table}. Experiments for Theorem \texttt{total-correctness-fib}.  We mark if the cluster contains (in addition to Theorem \texttt{total-correctness-fib}): 
  $g)$ Theorem \texttt{total-correctness-fact}, $h)$ Theorem \texttt{total-correctness-power}, $i)$ theorems related to Step 3 of the methodology.
  The size of the data set is $996$ theorems.}}\label{tab:compare-big}
	\vspace{.5cm}
 \end{table}

From Table~\ref{tab:compare-big}, we can draw the following conclusions:

\begin{itemize}
 \item Using K-means algorithm and $5$ as granularity value, the clusters are small and precise. Therefore, they produce the best input for the lemma analogy tool. 
 \item The method scale when we increase the size of the libraries (specially using K-means algorithm and $5$ as granularity value). 
 This can be seen from top tables of Tables~\ref{tab:compare} and~\ref{tab:compare-big} where the size and elements of clusters are similar.
 \item The clusters for terms with complex term-structures (for instance, Theorems \verb"program-correct-fib" and \verb"total-correctness-fib") are smaller than the ones for 
 simpler terms (see the sizes of the clusters of Table~\ref{tab:compare-big}). This indicates that the more complex the term-structure, the more precise the clusters.
\end{itemize}

\end{document}